\documentclass[twocolumn]{svjour3}          

\smartqed  
\usepackage{tabularx} 
\usepackage{amsmath}  
\usepackage{amssymb}
\usepackage{bm}
\usepackage{caption}
\usepackage{float}
\usepackage{graphicx} 
\usepackage{gensymb}
\usepackage[numbers]{natbib}
\usepackage[final]{hyperref} 

\begin{document}

\title{Behavior of confined granular beds under cyclic thermal loading}

\author{Pavel S. Iliev $^1$ \and
Elena Giacomazzi $^1$ \and
Falk K. Wittel $^1$ \and
Miller Mendoza $^{1,2}$ \and
Andreas Haselbacher $^3$\and
Hans J. Herrmann $^{1,4}$
}

\institute{
Pavel S. Iliev \\
\email{ilievp@ethz.ch} \\
Falk K. Wittel \\
\email{fwittel@ethz.ch} \\
$^1$Computational Physics for Engineering Materials, IfB, ETH Zurich, Stefano-Franscini-Platz 3, 8093 Zurich, Switzerland\\
$^2$ Simulation of Physical Systems Group, Universidad Nacional de Colombia, Departamento de Fisica, Cra 30 $\#$ 45-03, Ed. 404, Of. 348, Bogota D.C., Colombia\\
$^3$ Energy Science Center, ETH Zurich, Sonneggstrasse 28, 8092 Zurich, Switzerland\\
$^4$ PMMH, ESPCI, 7 Quai St. Bernard, 75005 Paris, France
}

\maketitle

\begin{abstract}
We investigate the mechanical behavior of a confined granular packing of irregular polyhedral particles under repeated heating and cooling cycles by means of numerical simulations with the Non-Smooth Contact Dynamics method. Assuming a homogeneous temperature distribution as well as constant temperature rate, we study the effect of the container shape, and coefficients of thermal expansions on the pressure buildup at the confining walls and the density evolution. We observe that small changes in the opening angle of the confinement can lead to a drastic peak pressure reduction. Furthermore, the displacement fields over several thermal cycles are obtained and we discover the formation of convection cells inside the granular material having the shape of a torus. The root mean square of the vorticity is then calculated from the displacement fields and a quadratic dependency on the ratio of thermal expansion coefficients is established.
\end{abstract}

\section{Introduction}\label{intro}
Heating-cooling cycles of a granular bed cause individual particles to expand and contract, leading to compaction and dilation of the granular system, which in turn causes thermal ratcheting when the coefficients of thermal expansion of the particles and the confining container are different. Plastic deformation of the granular material can result in large stresses in the system, which can lead to the fragmentation of particles as well as damaging of the container. A practical application where heating-cooling cycles of granular beds are important is packed-bed thermal-energy storage (TES)~\cite{ZanganehEtAl2012,Esence2017}. Their simplicity, high efficiency, and low cost make them attractive for use in concentrated solar power plants to reduce fluctuations in the produced energy and to allow for potentially continuous operation. Different material combinations for charge and confinement are being considered for TES. Natural rocks are of particular interest due to their abundance and low cost~\cite{BecattiniEtAl2017}. The thermo-mechanical behavior of granular beds in response to heating-cooling cycles is not fully understood and requires further investigation. In principle, local differences in thermal expansion will redistribute the particles, resulting in porosity variations that influence the flow of the heat-transfer fluid (HTF), leading to changes in the temperature fields, causing further particle redistributions.

In recent years, several studies on cyclic thermal loading of packed beds have been performed using Discrete Element Method (DEM) simulations with soft particles. Drei{\ss}igacker \textit{et al.}~\cite{Dreissigacker2010, Dreissigacker2013} presented a DEM model coupled to a simple model for the temperature fields of the particles and the HTF and applied it to 2D packings of discs in a rigid rectangular confinement. They studied the changes of the bed height and the force exerted on the vertical walls after one cycle. Sassine \textit{et al.}~\cite{Sassine2018} considered a 3D system with spherical particles and examined the pressure exerted on a rigid cylindrical container and the particle displacements. They investigated both homogeneous and moving linear temperature fields. A similar model was studied by Zhao \textit{et al.}~\cite{Zhao2016}, with a focus on entropy and statistical properties of the coordination number and contact forces in sphere packings. 

Until now, no numerical study appears to have been performed with irregular particles, which can drastically increase interlocking between particles. Moreover, thermal expansion of the container has also been neglected in previous studies. DEM simulations with polyhedral particles are well suited for the realistic representation of irregular granular materials such as rocks. We employ the Non-Smooth Contact Dynamics (NSCD) method~\cite{moreau1993,ContactDynamicsForBeginners} with polyhedral particles to study the thermal cycling of an initially loosely packed bed in a truncated conical container, see Fig.~\ref{fig:particle_container}\textbf{(a)}. Such containers have been suggested for packed-bed TES based on heuristic arguments about reducing the forces on particles~\cite{ZanganehEtAl2012}. In this work, we quantify the effect of the cone opening angle on the granular mechanics and we investigate how the stress buildup depends on the ratio between the thermal-expansion coefficients of particles and the container for the purely cylindrical geometry. Macroscopic quantities, such as the pressure acting on the container walls and the change in total volume are monitored. The compaction of the granular medium is obtained from the change in the total volume, which provides valuable insights into ratcheting effects~\cite{Marroquin2004,Farhadi2015} and the dilatancy of the granular material. We also examine microscopic properties like displacement fields and the resulting vorticity fields, which allows us to study the mechanisms behind the accumulation of plastic deformations in the bed material. The formation of convection cells is observed and the vorticity field is studied as a function of the cone opening angle and the ratio of the thermal-expansion coefficients.
\section{Materials and Methods}\label{matmed}
\begin{figure*}
\begin{center}
  \includegraphics[width=0.9\textwidth]{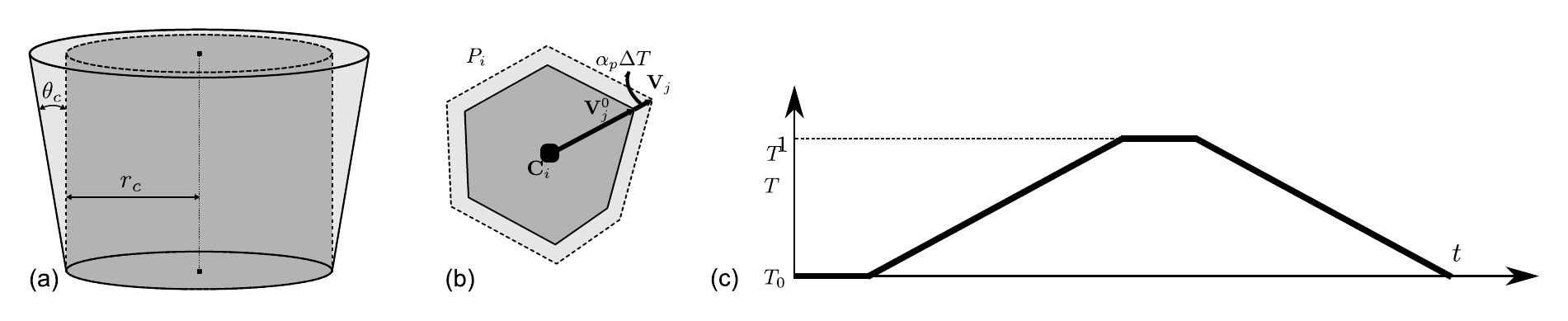}
\end{center}
\caption{Schematic illustration of \textbf{(a)} the container geometry defined by the radius $r_{c}$ and the opening angle $\theta_{c}$, \textbf{(b)} the thermal expansion of a single particle $P_{i}$, and \textbf{(c)} the temperature change during one thermal cycle.}\label{fig:particle_container}      
\end{figure*}

The packed bed is contained by a truncated cone with an opening angle $\theta_{c}$ of $0^\circ$ to $8^\circ$, see Fig.~\ref{fig:particle_container}\textbf{(a)}. The container is discretized by 16 planar segments and considered to expand with a thermal-expansion coefficient $\alpha_{c}$. The bed consists of $N_{p}$ convex polyhedral particles with expansion coefficient $\alpha_{p}$. The polyhedra are defined by their vertices and a list of faces. Following Refs. \cite{Iliev2018, Iliev2019}, we impose disorder and asymmetry by placing vertices randomly on the surface of an ellipsoid with normalized axes $a_{e}=1$, $b_{e}=0.95$, and $c_{e}=0.9$. The interaction between the particles is modelled by the NSCD method, based on a volume-exclusion constraint and Coulomb's friction law without regularization. This makes the method particularly well-suited for the modelling of dense packings of rigid, frictional particles with long-lasting contacts. Because of the discontinuous nature of the contact laws for NSCD we employ an implicit scheme for the integration of the equations of motion,
\begin{equation} \label{cd_eq}
m_{i} \frac{d}{dt} \vec{v}_{i} = \vec{F}_{i} \quad \text{and} \quad
\underline{\mathbf{I}}_{i} \frac{d}{dt} \bm{\omega}_{i} = \vec{T}_{i},
\end{equation}
where $m_{i}$ and $\underline{\mathbf{I}}_{i}$ are the mass and moment of inertia tensor, and $\vec{v}_{i}$ and $\bm{\omega}_{i}$ are the translational and angular velocities of particle $i$, and $\vec{F}_{i}$ and $\vec{T}_{i}$ are forces and torques acting on particle $i$. The force $\vec{F}_{i}$ is the sum of all contact forces and external forces, which we denote by $\vec{F}_{i}^{\text{cont}}$ and $\vec{F}_{i}^{\text{ext}}$ respectively. At each time step, $\vec{F}_{i}^{\text{cont}}$ is calculated with an iterative Gauss-Seidel algorithm until a global convergence criterion is fulfilled. The distances and normal vectors between two contacting particles are calculated by the Common Plane (CP) method~\cite{Cundall1988,Nezami2004,Nezami2006}. As illustrated in Fig.~\ref{fig:particle_container}\textbf{(b)}, particles expand isotropically, which is achieved by scaling the vector $\vec{V}_{j}$ connecting the $j$-th vertex to the particle's center of mass $\vec{C}_{i}$,
\begin{equation} \label{thermal_part_eq}
\vec{V}_{j} = \vec{V}^{0}_{j} \left( 1 + \alpha_{p} \Delta T^{t}_{t_0} \right), \quad \forall j \in N^{V}_{i},
\end{equation}
where $\vec{V}^{0}_{j}$ is the vector $\vec{V}_{j}$ at time $t_{0}$, $\Delta T^{t}_{t_0}$ is the temperature change and $N^{V}_{i}$ is the number of vertices of the polyhedral particle $P_{i}$. Since particles are ideally rigid in the NSCD method, the contact force is insensitive to overlaps. To capture the repulsive force of two particles overlapping from the thermal expansion, the signed distance $g$ between the closest points of approach for the contact between two particles is calculated according to $g = \max(1, p^\text{gap})g$, where $p^\text{gap}$ is a positive constant that is kept small to avoid a buildup of kinetic energy in the system~\cite{ContactDynamicsForBeginners}. A linear expansion is also imposed on the container.

For simplicity, the thermal field in the bed is considered to be homogeneous. Non-homogeneous thermal fields will be studied in future work. A thermal cycle is defined by a linear increase from the temperature $T_{0}$ to the temperature $T_{1}$, a relaxation at constant temperature, followed by a linear decrease to $T_{0}$, and another relaxation, see Fig.~\ref{fig:particle_container}\textbf{(c)}. Several thermal cycles were performed in the simulations. To gain a deeper understanding of the system behavior and to reduce the effect of random fluctuations on the results, we perform multiple realizations for each set of parameters. 

Typical simulations parameters are listed in Table~\ref{table:param}.  The opening angle $\theta_{c}$ and the coefficient of thermal expansion of the confinement $\alpha_{c}$ were varied while keeping $\alpha_{p}$ constant. Due to the axial symmetry of the confinement we perform an averaging around the central axis using a grid. This results in a mapping to cylindrical coordinates, where an averaging over the polar angle is performed. The dimensionality of the system is thereby reduced, allowing the above-mentioned properties to be represented as functions of the radial distance from the central axis and the vertical distance.
\begin{table}
\centering
\caption{Summary of typical values of the dimensionless simulation parameters.}
\label{table:param}
\begin{tabular}{p{4.2cm}|p{1.4cm}|p{1.4cm}}								  
\textbf{Description}       		& \textbf{Name}     & \textbf{Value}        \\
\hline
Number of particles       		& $N_{p}$           & $2000$        \\
Container radius       		    & $r_{c}$           & $8.5$         \\
Minimum particle radius		    & $r_{\min}$         & $0.5$         \\
Maximum particle radius         & $r_{\max}$         & $1.0$         \\
Lowest temperature         		& $T_{0}$           & $0$           \\
Highest temperature       		& $T_{1}$           & $100$         \\
Thermal expansion coefficient   & $\alpha_{p}$      & $10^{-4}$     \\
Thermal expansion coefficient   & $\alpha_{c}$      & $10^{-3}$     \\
\end{tabular}
\end{table}
\section{Results}\label{res}
\begin{figure*}[hbt]
\begin{center}
  \includegraphics[width=1\textwidth]{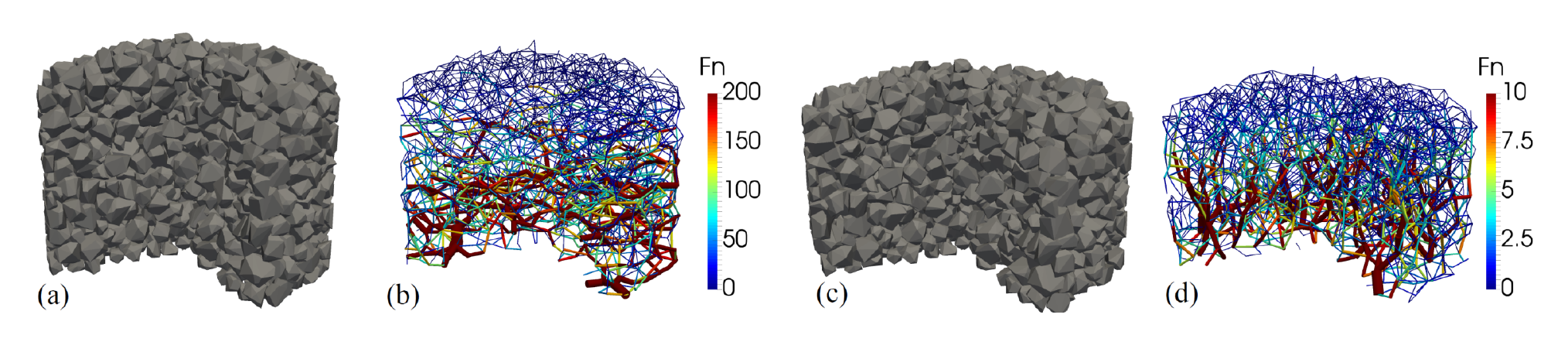}
\end{center}
\caption{Snapshots from a single realization for $\theta_{c} = 0 ^\circ$ and $\alpha_{c} / \alpha_{p} = 10$ after 20 thermal cycles. \textbf{(a)} The particles at the end of a cooling stage and \textbf{(b)} the corresponding normal contact force network. \textbf{(c)} The particles at the end of a heating stage and \textbf{(d)} the corresponding normal contact force network.}\label{fig:sim_snap}      
\end{figure*}

A single realization of the system is shown in Fig.~\ref{fig:sim_snap} \textbf{(a)} and \textbf{(c)} for $\theta_{c} = 0 ^\circ$ and $\alpha_{c} / \alpha_{p} = 10$ at the ends of a cooling stage and a heating stage, respectively. During the cooling stage, strong force chains form in the lateral direction in the lower part of the container because of the shrinkage of the confinement. This leads to a stress buildup at the container walls as can be seen in Fig.~\ref{fig:sim_snap}\textbf{(b)}. To minimize the potential energy of the system, particles rotate and rearrange to find a more optimal configuration, which results in an upward motion at the center of the container. During the heating stage, confining forces weaken, see Fig.~\ref{fig:sim_snap}\textbf{(d)}. Near the container walls, gaps appear that are filled by particles, which on average leads to a downward motion of the particles close to the walls. 

\begin{figure*}
\begin{center}
  \includegraphics[width=1\textwidth]{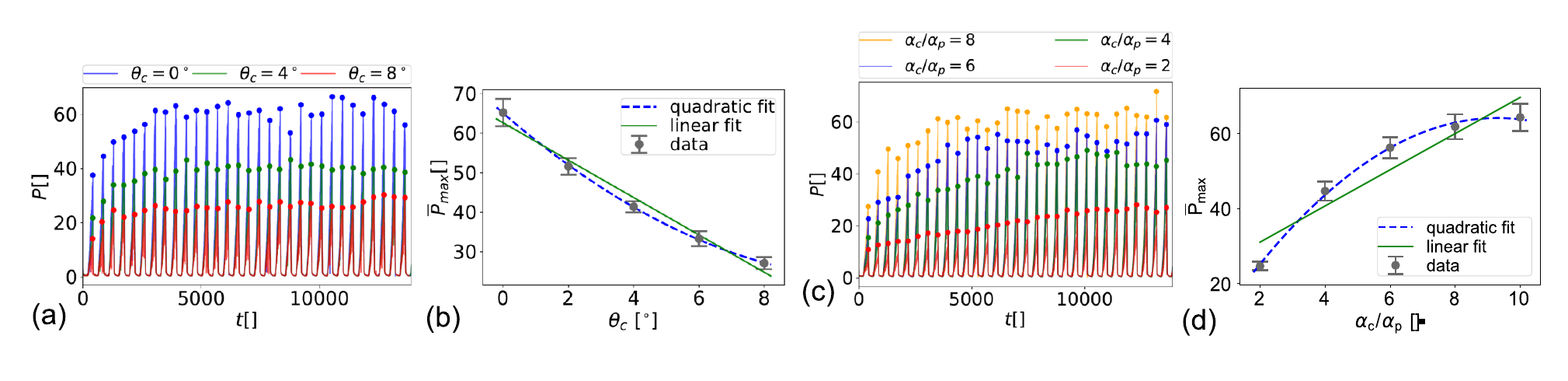}
\end{center}
\caption{The pressure on the walls as a function of \textbf{(a)} the angle $\theta_{c}$ and \textbf{(c)} the ratio $\alpha_{c} / \alpha_{p}$. The peak value for each cycle is denoted by a dot. The dependence of the mean peak pressure $\overline{P}_{\max}$ during the steady state on \textbf{(b)} the angle $\theta_{c}$ and \textbf{(d)} the ratio $\alpha_{c} / \alpha_{p}$. Each data point represents an average over 5 realizations and the error bars indicate one standard deviation. }\label{fig:pram_study}      
\end{figure*}

Since high contact forces can fracture particles, which reduces the efficiency and lifetime of a TES, we are interested in the time evolution of the pressure $P$ acting on the container walls. When monitoring this pressure, two regimes appear for all parameters. In the first regime, i.e., during the first cycles, the pressure increases. In the second regime, a steady state is reached where the pressure fluctuates around a relatively constant value, see Figs.~\ref{fig:pram_study}\textbf{(a)} and \textbf{(c)}. This behavior was also observed by Sassine \textit{et al.} \cite{Sassine2018}. The first regime is due to compaction of the granular material, since pressure increases as the total packing density increases. After the maximum density is reached, we also observe a saturation in the pressure peaks. We discuss first the influence of the opening angle $\theta_{c}$. From Fig.~\ref{fig:pram_study}\textbf{(a)}, we observe that the steady state is reached quicker with increasing $\theta_{c}$ and that the pressure peaks at the steady state decrease with increasing $\theta_{c}$. The dependence of the mean peak pressure $\overline{P}_{\max}$ on $\theta_{c}$ is shown on Fig.~\ref{fig:pram_study}\textbf{(b)} and appears to be approximated well by a quadratic function, but does not deviate strongly from a linear relation either. Remarkably, a conical container with an opening angle of only $6^{\circ}$ halves the peak pressure compared to a cylindrical container. Next, we discuss the dependency of $\overline{P}_{\max}$ on $\alpha_{c} / \alpha_{p}$ as shown in Figs.~\ref{fig:pram_study}\textbf{(c)} and \textbf{(d)}. It is evident that $\alpha_{c} / \alpha_{p}$ introduces a time scale that determines the rate of peak pressure increase as a function of the number of cycles. The ratio $\alpha_{c} / \alpha_{p}$ is equivalent in this configuration to the energy input to the system as it determines the amount of deformation of the container when the temperatures $T_{0}$ and $T_{1}$ are fixed. As we see from Fig.~\ref{fig:pram_study}\textbf{(d)}, the peak pressure increases with increasing $\alpha_{c} / \alpha_{p}$ as expected and is fitted well by a quadratic function.

\begin{figure}
\begin{center}
  \includegraphics[width=0.48\textwidth]{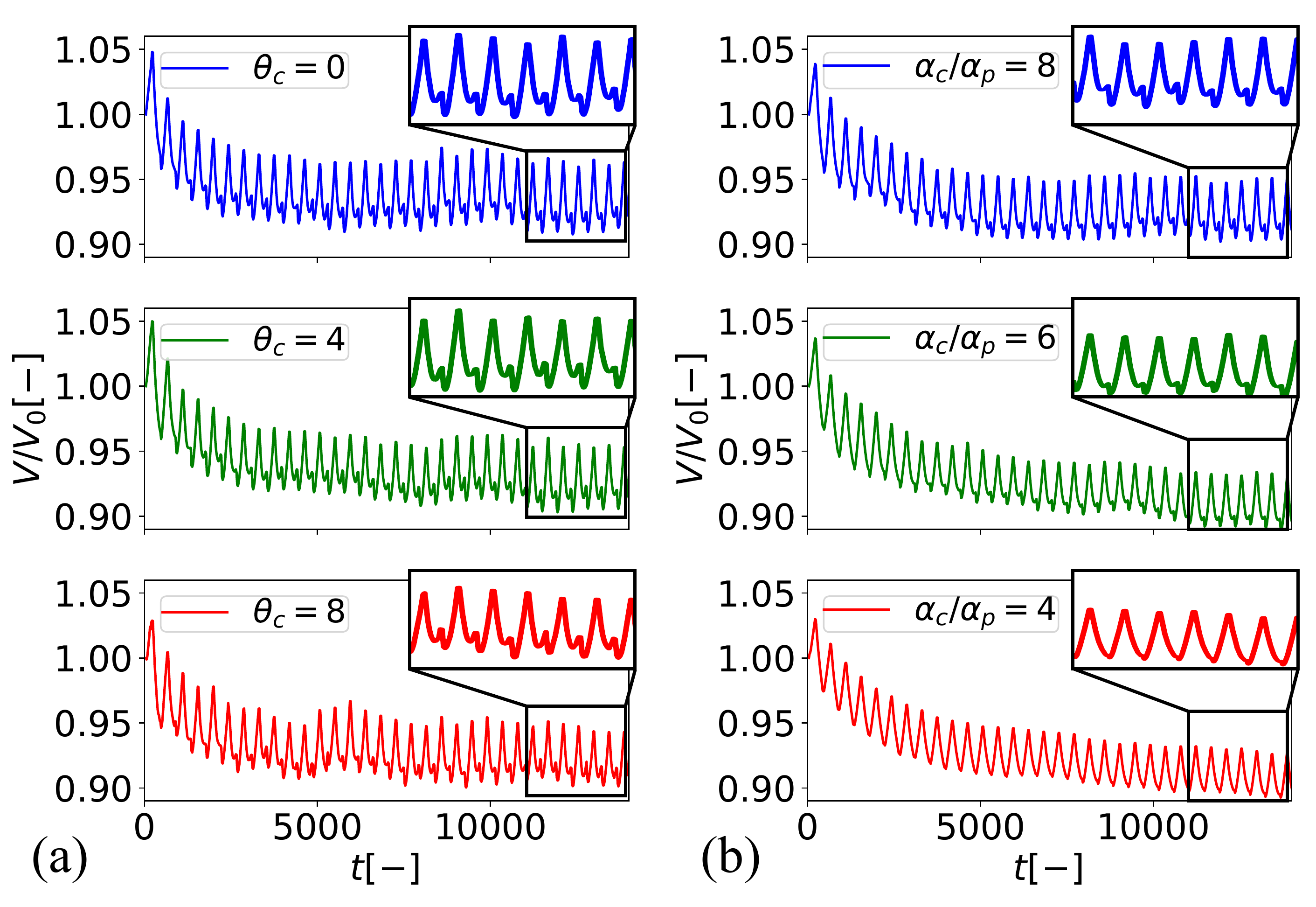}
\end{center}
\caption{Volume $V / V_{0}$ as a function of \textbf{(a)} $\theta_{c}$ with fixed $\alpha_{c} / \alpha_{p} = 10$ and \textbf{(b)} $\alpha_{c} / \alpha_{p}$ with fixed $\theta_{c} = 0 ^{\circ}$.}\label{fig:volume_change}       
\end{figure}

Understanding the compaction and dilatation mechanisms of the particles in the container is important because changes in porosity affect the flow of the HTF which in turn affects the TES efficiency. Therefore, we study the total volume change, defined as the total volume of the granular material $V$ normalized by the initial volume $V_{0}$ before thermal cycling. At each time step, $V$ is calculated from the convex hull of the vertices of all particles. The time evolution of $V / V_{0}$ is shown in Fig.~\ref{fig:volume_change} as a function of $\theta_{c}$ and $\alpha_{c} / \alpha_{p}$. We observe a decrease in the volume, equivalent to an increase in the packing density, for the first few cycles during the compaction regime, which is followed by a saturation that corresponds to the steady state observed for the pressure. Closer investigation of the oscillatory nature of $V / V_{0}$ indicates small increases during the cooling at the steady state. These increases are caused by dilatation of the granular material in the high-density regime. Although dilatation is present for all angles $\theta_{c}$, see Fig.~\ref{fig:volume_change}\textbf{(a)}, we observe that the magnitude of dilatation decreases with decreasing $\alpha_{c}$. 

\begin{figure}
\begin{center}
  \includegraphics[width=0.48\textwidth]{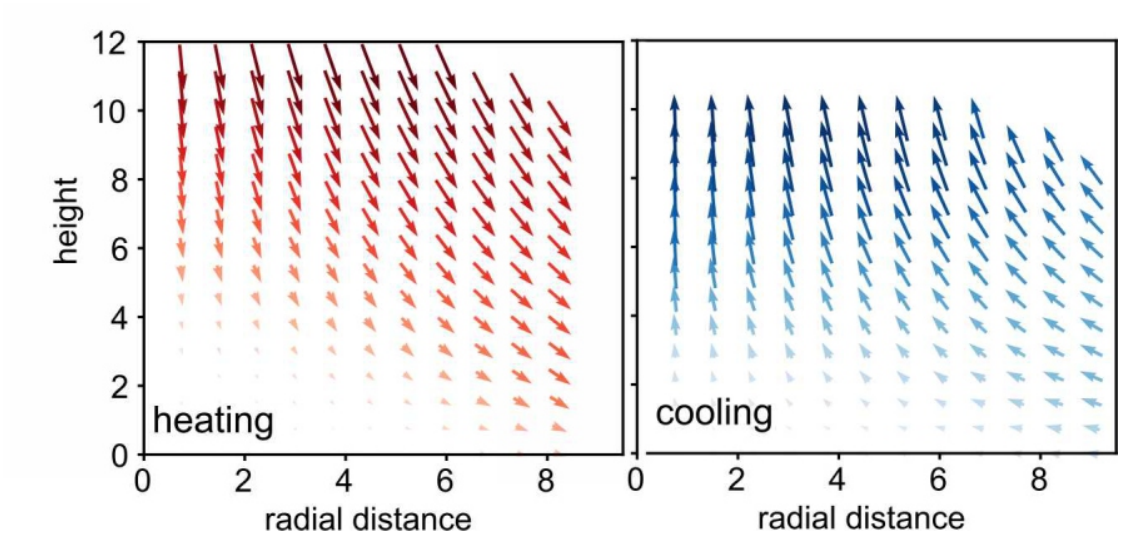}
\end{center}
\caption{Displacement fields during a heating stage, and a cooling stage for $\theta_{c} = 0 ^\circ$ and $\alpha_{c} / \alpha_{p} = 10$.}\label{fig:displ}    
\end{figure}

To gain a more complete understanding of the material deformation, we turn to the displacement fields. First we show the displacement fields during thermal loading and unloading for $\theta_{c} = 0 ^\circ$ and $\alpha_{c} / \alpha_{p} = 10$, see Fig.~\ref{fig:displ}. As the system expands during the temperature increase, particles can fill gaps that open near the walls, resulting in a motion from the central axis outwards and downwards. A conical zone at the bottom of the system with almost no displacements can also be seen, indicating the formation of shear bands. Conversely, as the system contracts during the temperature decrease, the shrinking walls of the container push the particles inwards and upwards. Although the two motions appear to be simple, we will show that the cumulative effect is non-trivial.

\begin{figure}
\begin{center}
  \includegraphics[width=0.48\textwidth]{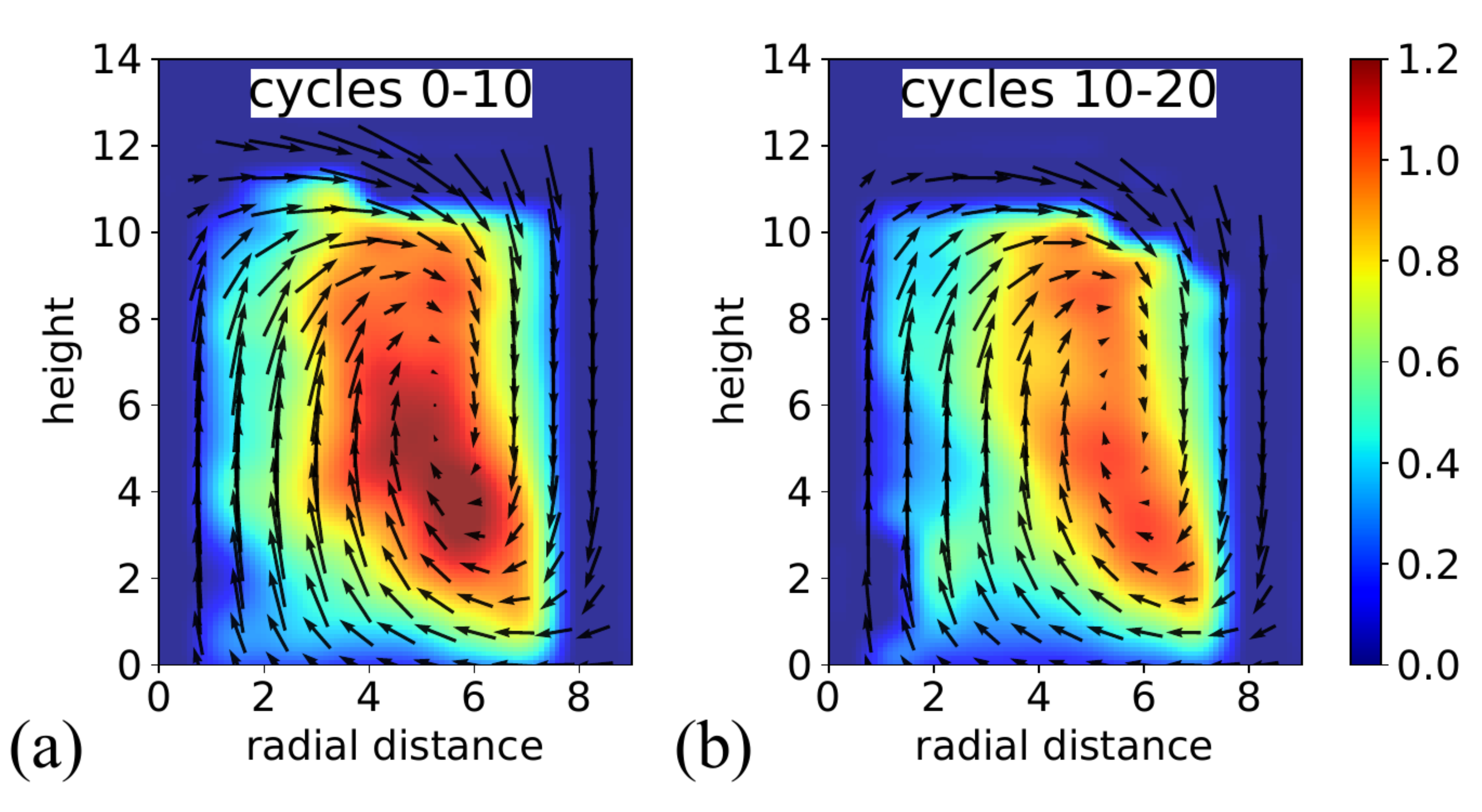}
\end{center}
\caption{
Accumulated displacement and vorticity fields for \textbf{(a)} thermal cycles 0 to 10, and \textbf{(b)} thermal cycles 10 to 20 for $\theta_{c} = 0 ^\circ$ and $\alpha_{c} / \alpha_{p} = 10$. The colors represent the vorticity magnitude.
}
\label{fig:displ_vort}      
\end{figure}

Investigating the nature of granular convection is of intrinsic scientific interest and may be important for the improvement of TES. The main mechanism responsible for the plastic deformation over the course of many thermal cycles becomes clear if accumulated displacements are monitored and visualized. In Fig.~\ref{fig:displ_vort}, we show the total accumulated displacement fields over 10 cycles for the first and second 10 cycles for $\theta_{c} = 0 ^\circ$ and $\alpha_{c} / \alpha_{p} = 10$, respectively. One immediately notices the formation of convection cells in the bed accounting for the plastic deformation due to thermal cycling, or more precisely, the ratcheting effects due to the voids appearing with the deformation of the confinement. Because of rotational symmetry, the convection cells take the form of a toroidal vortex. The convection cells were observed for all combinations of system parameters investigated in this study. Convection cells in granular systems have been observed and investigated repeatedly, such as in vertically vibrated beds ~\cite{Gallas1992,Yidan1997,Zhang2014}, horizontally vibrated beds ~\cite{Liffman1997,Hsiau2002}, vibrated hopper flow ~\cite{Wassgren2002}, and shear flow in the presence of a temperature gradient ~\cite{Rognon2010}. To quantify the magnitude of the rotational motion, we calculate the curl $w = \partial u_{z}/\partial r - \partial u_{r}/\partial z$ from the total displacement field $\vec{u}=(u_{r}, u_{z})$ using the scheme proposed in \cite{Sumesh2013}. The vorticity fields are shown in Fig.~\ref{fig:displ_vort} superimposed on the displacement fields. To remove boundary effects, the curl is calculated inside the displacement field and all boundary cells are neglected. The magnitude of the vorticity in the first 10 cycles is higher than in the second 10 cycles because when the system has lower density, the particles can move more freely faster. The convective motion could be harmful since the particles can wear and fragment, causing accumulation of dust and fragments near the bottom of a TES, which may lead to non-uniform flow of the HTF, increased pumping requirements, and reduced efficiency.

\begin{figure}
\begin{center}
  \includegraphics[width=0.48\textwidth]{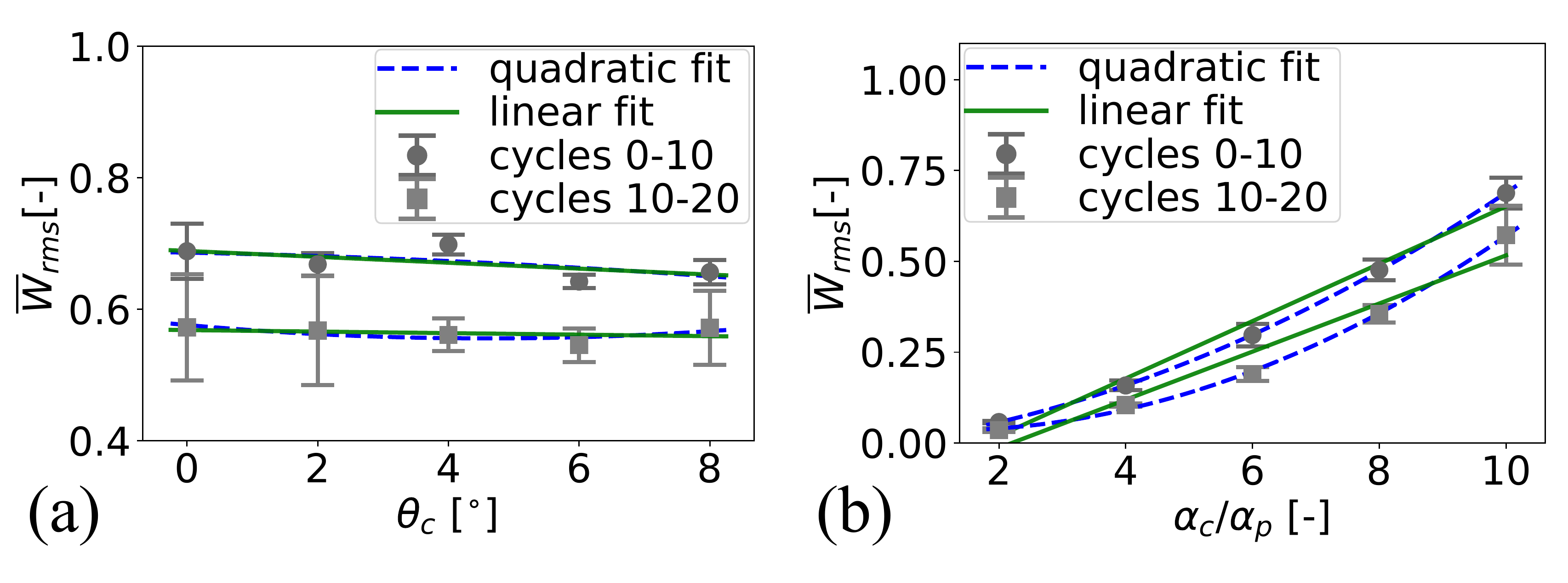}
\end{center}
\caption{Average root mean square vorticity $\overline{W}_{rms}$ as a function of \textbf{(a)} $\theta_{c}$ for $\alpha_{c}/\alpha_{p} = 10$ and \textbf{(b)} $\alpha_{c}/\alpha_{p}$ for $\theta_{c} = 0 ^\circ$.}\label{fig:vort_study}      
\end{figure}

Lastly, to quantify the overall magnitude of the vorticity over several cycles, we calculate the root mean square vorticity (RMS) $\overline{W}_{rms}$, averaged over five realizations for various values of $\theta_{c}$ and $\alpha_{c}/\alpha_{p}$, see Fig.~\ref{fig:vort_study}. Within the error bars, $\overline{W}_{rms}$ is practically independent of $\theta_{c}$ at any stage of the thermal cycling as we see from the approximately horizontal linear fits in Fig.~\ref{fig:vort_study}\textbf{(a)}, however, the magnitudes are larger at the beginning stages of the thermal loading. Because the energy scales with the square of the strain, the dependence of $\alpha_{c} / \alpha_{p}$ on $\overline{W}_{rms}$ appears to be well fitted by a quadratic function and depends on the stage of the system evolution, see Fig.~\ref{fig:vort_study}\textbf{(b)}. Overall, we observe that the magnitude of the vorticity increases monotonically with $\alpha_{c}$ and the deviation from the linear approximation is stronger in the steady state, i.e., after the first 10 cycles. 
\section{Conclusion and Outlook}
We developed a DEM of a granular material consisting of irregular polyhedral particles that allows for deformation due to temperature change and employed it to study the mechanical behavior of a confined system under heating/cooling cycles. We investigated a configuration resembling a TES system assuming for simplicity a homogeneous temperature distribution and linear temporal increases/decreases in the temperature as well as a linear law for the thermal expansion of the particles and the container. 

An uncompacted regime and a saturated, compacted regime were identified. We established that the peak pressure at the steady state does not deviate strongly from a linear dependence on the opening angle of the container wall. The dependence of the pressure on the thermal coefficient of the confinement, however, appears to follow a quadratic relation. We further investigated the time evolution of the total volume change, from which we concluded that in the steady state, the material reaches nearly the same peak density for all investigated parameters, but differences in the dilatation magnitude appear and almost no dilatation is visible for small values of $\alpha_{c}$. 

Furthermore, for any combination of the investigated parameters, we observe the formation of convection cells in the form of a toroidal vortex over the course of many cycles. This form of granular flow in the confinement can lead to accelerated damage of the material and also to size segregation and heterogeneous packing density, which can be harmful for the longevity of such systems. Moreover, we calculated the magnitude of the vorticity fields from the cumulative displacements of the particles over several cycles. Although an increase in $\theta_{c}$ led to a strong decrease in pressure, a change in the RMS vorticity was not observed. But the dependency of $\alpha_{c}$ on $\overline{W}_{rms}$ appears to follow a quadratic relation in both the uncompacted and the compacted regimes. 

Further investigations should include more realistic temperature distributions. This would require a more accurate representation of the container walls as they will also deform according to this temperature distribution. Fluid-particle coupling as well as multifield simulations with heat exchange are also important topics. Another question to be addressed is whether an increase of the system size, i.e., the number of particles,  will lead to different behavior, such as the formation of additional vortices. Finally, the cases $\alpha_{c}/\alpha_{p} < 1$ as well as $\theta_{c} > 8 ^\circ$ would be interesting to consider and determine whether there is a set of parameters for which the convection changes  direction. \\
\begin{acknowledgements}
We acknowledge financial support from the ETH Research Grant ETHIIRA Grant No. ETH-04 14-2 as well as from CAPES and FUNCAP and by the Swiss Commission for Technology and Innovation through the Swiss Competence Center for Energy Research on Heat and Electricity Storage. We also thank D. Rubis and P. Hilger for their help with the preliminary simulations. \\
\end{acknowledgements}

\end{document}